\documentclass[12pt]{article}
\usepackage{amsthm,amsmath,latexsym,amssymb,amsfonts,amscd}
\usepackage{cite,graphics,lscape,fancyhdr,array,stmaryrd,euscript}
\usepackage{ifthen,empheq,slashed}
\usepackage{sidecap}
\usepackage{ifpdf}
\usepackage{verbatim}
\usepackage{color}
\usepackage{relsize}
\usepackage[utf8]{inputenc}

\numberwithin{equation}{section}
\newcommand{\p}[1]{(\ref{#1})}
\newcommand{\be}{\begin{equation}}
\newcommand{\bea}{\begin{eqnarray}}
\newcommand{\ee}{\end{equation}}
\newcommand{\eea}{\end{eqnarray}}

\newcommand{\ga}{\alpha}
\newcommand{\gb}{\beta}

\newcommand{\pl}{\partial}

\newcommand{\bep}{\begin{picture}}
\newcommand{\eep}{\end{picture}}

\newcounter{YoungHeight}\newcounter{YoungWidth}

\newcounter{Mul1}\newcounter{Mul2}\newcounter{Mul3}\newcounter{Mul4}
\newcounter{A0}\newcounter{A1}\newcounter{A2}
\newcounter{B3}
\newcounter{C3}\newcounter{C4}
\newcounter{D1}\newcounter{D2}\newcounter{D3}
\newcounter{T0}\newcounter{T1}

\newlength{\txtHShift}

\newlength{\txtWidth}

\newcommand{\HalfLength}[2]{\setcounter{Mul1}{#1}\setcounter{Mul2}{#1}\addtocounter{Mul1}{\value{Mul2}}\addtocounter{Mul1}{\value{Mul2}}%
\addtocounter{Mul1}{\value{Mul2}}\addtocounter{Mul1}{\value{Mul2}}\setcounter{#2}{\value{Mul1}}}

\newcommand{\Add}[3]{\setcounter{#1}{#2}\addtocounter{#1}{#3}}

\newcommand{\Length}[1]{#10}
\newcommand{\YoungScale}{}

\newcommand{\shiftedText}[2]{{\hspace{#1}#2}}
\newcommand{\calcHShift}[1]{\settowidth{\txtWidth}{#1}\setlength{\txtHShift}{-0.5\txtWidth}}

\newcommand{\TextTop}[3]{{\calcHShift{#1}\HalfLength{#2}{T0}\Add{T1}{\Length{#3}}{-9}\put(\value{T0},\value{T1}){\shiftedText{\txtHShift}{#1}}}}

\newcommand{\BlockA}[2]{{\YoungScale\bep(\Length{#1},\Length{#2}){\Add{A1}{#1}{1}\Add{A2}{#2}{1}}%
\multiput(0,0)(10,0){\value{A1}}{\line(0,1){\Length{#2}}}\multiput(0,0)(0,10){\value{A2}}{\line(1,0){\Length{#1}}}%
\setcounter{YoungHeight}{\Length{#2}}\setcounter{YoungWidth}{\Length{#1}}\eep}}

\newcommand{\RectT}[3]{\bep(\Length{#1},\Length{#2})\put(0,0){\line(1,0){\Length{#1}}}\put(0,0){\line(0,1){\Length{#2}}}%
\put(\Length{#1},\Length{#2}){\line(-1,0){\Length{#1}}}\put(\Length{#1},\Length{#2}){\line(0,-1){\Length{#2}}}#3{#1}{#2}\eep}

\newcommand{\RectARow}[2]{{\bep(\Length{#1},10)\put(0,0){\RectT{#1}{1}{\TextTop{#2}}}\eep}}

\newcommand{\YoungBB}{\BlockA{2}{2}}

\newcommand{\BlockApar}[2]{\parbox{\Length{#1}pt}{\YoungScale\bep(\Length{#1},\Length{#2}){\Add{A1}{#1}{1}\Add{A2}{#2}{1}}%
\multiput(0,0)(10,0){\value{A1}}{\line(0,1){\Length{#2}}}\multiput(0,0)(0,10){\value{A2}}{\line(1,0){\Length{#1}}}%
\setcounter{YoungHeight}{\Length{#2}}\setcounter{YoungWidth}{\Length{#1}}\eep}}

\newcommand{\BlockBpar}[4]{\parbox{\Length{#1}pt}{\YoungScale\Add{B3}{\Length{#2}}{\Length{#4}}%
\bep(\Length{#1},\value{B3})\put(0,\Length{#4}){\BlockA{#1}{#2}}%
\put(0,0){\BlockA{#3}{#4}}\setcounter{YoungHeight}{\value{B3}}\setcounter{YoungWidth}{\Length{#1}}\eep}}

\newcommand{\YoungpBB}{\BlockApar{2}{2}}

\newcommand{\YoungpDB}{\BlockBpar{4}{1}{2}{1}}

\usepackage{hyperref}

\begin{document}
\topmargin -1cm \oddsidemargin=0.25cm\evensidemargin=0.25cm
\textwidth 18cm
\setcounter{page}0
\renewcommand{\thefootnote}{\fnsymbol{footnote}}

\begin{titlepage}
\begin{flushright}
LMU - ASC 40/16 \\
\end{flushright}
\vskip .7in
\begin{center}
{\LARGE \bf   Correlation Functions of
$Sp(2n)$ Invariant Higher-Spin Systems}
\vskip .6in {\Large Evgeny Skvortsov$^{a,b}$\footnote{e-mail: {\tt  evgeny.skvortsov@physik.uni-muenchen.de}},
Dmitri Sorokin$^c$\footnote{e-mail: {\tt  dmitri.sorokin@pd.infn.it }} and
 Mirian Tsulaia$^d$\footnote{e-mail: {\tt  mirian.tsulaia@uwa.edu.au}  }}
 \vskip .4in {$^a$ \it Arnold Sommerfeld Center for Theoretical Physics,
 Ludwig-Maximilians University Munich,
Theresienstr. 37, D-80333
   Munich, Germany} \\
\vskip .2in {$^b$ \it Lebedev Institute of Physics, Leninsky ave 53, 119991, Moscow, Russia} \\
\vskip .2in {$^c$ \it INFN, Sezione di Padova, via F. Marzolo 8, 35131 Padova, Italia} \\
\vskip .2in { $^d$ \it School of Physics $M013$,
The University of
Western Australia, 35 Stirling Highway,
Crawley, Perth, WA 6009, Australia}\\
\vskip .8in

\begin{abstract}
We study the general structure of correlation functions in an $Sp(2n)$-invariant formulation of systems of an infinite number of higher-spin fields. For $n=4,8$ and 16 these systems comprise the conformal higher-spin fields in space-time dimensions $D=4,6$ and 10, respectively, while when $n=2$, one deals with conventional $D=3$ conformal field theories of scalars and spinors. We show that for $n>2$ the $Sp(2n)$ symmetry and current conservation makes the 3-point correlators of two (rank-one or rank-two) conserved currents with a scalar operator be that of free theory.
This situation is analogous to the one in conventional conformal field theories, where conservation of  higher-spin currents implies that the theories are free.
\end{abstract}

\end{center}

\vfill

\end{titlepage}

\tableofcontents

\renewcommand{\thefootnote}{\arabic{footnote}}

\section{Introduction}

The $Sp(2n)$ invariant description of 
massless bosonic and fermionic higher-spin fields \cite{Fronsdal:1985pd,Bandos:1998vz,Bandos:1999qf,Vasiliev:2001dc,Vasiliev:2001zy,Vasiliev:2002fs,Didenko:2003aa,Plyushchay:2003gv,Gelfond:2003vh,Plyushchay:2003tj,Vasiliev:2003jc,Bandos:2004nn,Bandos:2005mb,Gelfond:2006be,Vasiliev:2007yc,Ivanov:2007vx,Gelfond:2008ur,Gelfond:2008td,Gelfond:2010pm,Bandos:2011wi,Florakis:2014kfa,Florakis:2014aaa,Fedoruk:2012ka,Gelfond:2015poa}
is an elegant geometrical approach to study higher-spin gauge theories.
The main feature of this approach is that the theory is
formulated in an extended space, sometimes called hyperspace, which is parametrized by the $n\times n$ matrix valued coordinates
$X^{\alpha \beta}= X^{\beta \alpha}$. These $\frac{n(n+1)}{2}$ coordinates
include, in addition to ordinary space-time coordinates $x^m$ parameterizing either a $D$-dimensional Minkowski space or an anti-de-Sitter space $(AdS_D)$, also $\left[\frac{n(n+1)}{2}-D\right]$ extra coordinates
$y$.
The fields depend on the both $x$- and $y$-coordinates
and obey free field equations \cite{Vasiliev:2001zy} which are invariant under the
transformations of the $Sp(2n)$ group. The
analysis of these field equations for different $n$ \cite{Vasiliev:2001zy,Bandos:2005mb}
shows that for $n=4,8$ and $16$ they generate the field equations and the Bianchi identities for an infinite set of free conformal higher--spin curvatures in space-times of dimension $D=4,6$ and $10$, respectively\footnote{The case of $n=2$  corresponds to conventional $D=3$ conformal theories of scalar and spin-half fields, with $Sp(4)\sim SO(2,3)$ being the $3D$ conformal group. }. The $D$-dimensional linearized higher-spin curvatures $R_{(s)}(x)$ are the components of a series expansion of hyperfields $\Phi(x,y)$ in powers of $y$, schematically $\Phi(x,y)=\sum_{s=0}^{\infty}R_{(s)}(x) y^s$. 
The main hyperfields are a scalar $b(X)$ and a fermion $f_\alpha(X)$ $(X=(x,y))$ transforming under the linear representation of $GL(n)\subset Sp(2n)$. We will somewhat loosely call  $f_\alpha(X)$ the spinor field since it contains half-integer spin curvatures in the corresponding $D$-dimensional space-time and at $y=0$  reduces to a spinor field.

The group $Sp(2n)$ is often referred to as a generalised
conformal group, since its structure closely reminds the structure of the conformal group and moreover
the $Sp(2n)$ group contains a conformal group as a subgroup.
This fact not only makes the $Sp(2n)$ formulation
relevant to the study of conformal properties of higher-spin fields 
but is also useful for better understanding the structure of $Sp(2n)$-invariant systems themselves, as we shall see below.

A natural question to ask is whether $Sp(2n)$-invariant higher-spin systems admit interactions.
An unsuccessful attempt to obtain interacting $Sp(2n)$-invariant models was undertaken in \cite{Bandos:2004nn} in the framework of a generalized supergravity in tensorial superspaces.

Recently, it was shown \cite{Gelfond:2015poa} that in $D=4$ models higher-spin current interactions necessarily break $Sp(8)$
group down to the four-dimensional conformal group $SU(2,2)$. However, this result, a priori, does not rule out other types of  $Sp(2n)$-interactions, e.g. of some order in higher-spin curvatures. Moreover, there might exist $Sp(2n)$-theories that do not have any Lagrangian/equations of motion description at all, which is the case for certain  conventional conformal field theories.

One way of approaching the interaction problem generically is to use an analogy with $D$-dimensional
conformal field theories. One can make a statement whether a generic $D$-dimensional conformal theory is free or interacting by looking at the structure of its correlation functions. In this way, for instance, in \cite{Maldacena:2011jn} it was shown that under certain assumptions about the content of $D=3$ conformal theory, the presence in the theory of a single conserved higher-spin current implies the existence of an infinite set of conserved higher-spin currents whose correlators with the stress tensor are those of a free conformal theory. 

In all the examples of correlators in the $Sp(2n)$-invariant theories considered so far \cite{Vasiliev:2003jc,Florakis:2014kfa,Florakis:2014aaa} generalized conformal weights of the fields in the correlators were not restricted to their canonical values thus, in principle, leaving  room for nontrivial interactions. In this paper we will show that for $n> 2$ already the presence in the $Sp(2n)$-invariant theories of a conserved generalized stress tensor and/or of a conserved current  associated with a rigid internal symmetry  (introduced in \cite{Vasiliev:2002fs}) makes their correlators with a scalar operator to be those of free theories. To arrive at this result, the correlation functions
were constructed solely under the requirement of their $Sp(2n)$ invariance and current conservation properties, without resorting to a specific form of the operators. The three-point functions turn out to have a  structure, whose generating functions were found earlier in \cite{Giombi:2010vg,Colombo:2012jx,Didenko:2012tv,Gelfond:2013xt,Didenko:2013bj} with the use of a different approach.

Therefore, one concludes that for $n>2$ the  rigid $Sp(2n)$ symmetry, together with the conservation requirements, turn out
to be too restrictive for the existence of the nontrivial interactions of higher-spin fields already
at the cubic level. This confirms and generalises the result obtained in \cite{Gelfond:2015poa}. In order to allow for nontrivial interactions of these systems, the rigid $Sp(2n)$ invariance should be broken down to an appropriate subgroup.

 Let us emphasise  that our results apply to the systems with rigid $Sp(2n)$
symmetry.
It would be very interesting to construct and study systems which posses a local $Sp(2n)$
invariance, and see if analogous obstructions for interactions apply also in these cases.

The  paper is organised as follows.
In Section \ref{MHSF} we introduce a set up for the rest of the paper. We review  $Sp(2n)$ transformations
of the scalar field $b(X)$ and the spinor field $f_\alpha(X)$ and their free field equations. We also introduce
a generalised current $J_{\alpha \beta}(X)$ and a stress tensor $T_{\alpha \beta \gamma \delta}(X)$, and  give their
$Sp(2n)$ transformations and conservation laws.

In Section \ref{mpcf} we  explain the general procedure of constructing the $Sp(2n)$-invariant correlation functions and derive two-point functions of  two  currents  and two stress tensors using the  requirement of $Sp(2n)$ invariance and the current conservation.

In Section \ref{STHPF} we derive three-point functions
which include two scalar or spinor fields and one current or stress tensor. The results obtained in this Section are completely analogous to those in ordinary $D$-dimensional CFTs in the sense that there is no restriction on the generalised conformal dimensions of the scalar and spinor operators.

In Section \ref{SOTPF} we derive three-point functions of a scalar operator with two conserved currents or stress tensors. Here the situation turns out to be different from the one
in  $D$-dimensional CFTs. Namely, apart from the $n=2$, $D=3$ case, the values of the conformal dimensions of the scalar operators of the $Sp(2n)$-invariant systems in these correlation functions are fixed.

In Section \ref{HSalg} we discuss the generic $Sp(2n)$-invariant structure of three-point correlation functions of higher-rank tensorial fields whose three building blocks are provided by basic two- and three-point correlators of bosonic and fermionic fields and their currents.

In Conclusion we discuss the obtained results and their implications. Some lengthy calculations are given in the Appendices.

\section{$Sp(2n)$-invariant systems}\label{MHSF}

\subsection{Scalar field $b(X)$}
The basic object in the $Sp(2n)$-invariant description of integer higher-spin fields is a hyperfield $b(X^{\mu \nu})$ (see \cite{Florakis:2014kfa} for a recent review).
The hyperspace coordinates $X^{\mu\nu}=X^{\nu\mu}$ and the field $b(X)$ transform under the $Sp(2n)$ transformations as follows
\begin{equation}\label{deltaX}
\delta X^{\mu\nu}=a^{\alpha\beta}+X^{\mu\rho}g_\rho{}^{\nu}+X^{\nu\rho}g_\rho{}^{\mu}-X^{\mu\rho}k_{\rho\lambda}X^{\lambda\nu}\,,
\end{equation}
\begin{equation}\label{sp8fbD}
\delta b(X)= -{\Large(}a^{\mu \nu} \partial_{\mu \nu} + \Delta\,(g_\mu{}^\mu-
 k_{\mu \nu}  X^{\mu \nu}) + 2 g_\nu{}^\mu X^{\nu \rho}\,
\partial_{\mu \rho} -
 k_{\mu \nu}  X^{\mu \rho} X^{\nu \lambda}\partial_{\rho \lambda}{\large)} b(X)\,,
\end{equation}
where  the parameter $a^{\mu \nu}= a^{\nu \mu}$
corresponds to the translations, the parameter $k_{\mu \nu} = k_{\nu \mu}$ corresponds to the generalised conformal boosts and the parameter
$g_\mu{}^\nu$ is that of $GL(n)$ transformations. The latter can be split into $l_{\mu}{}^{\nu}=g_{\mu}{}^{\nu}-\frac {1}{n} \delta_\mu^\nu\, g_{\rho}{}^{\rho}$ which parametrizes the $sl(n)$ subalgebra of $gl(n)$ and its trace $g_{\mu}{}^{\mu}$ which corresponds to the dilatations.
Together these transformations generate an $Sp(2n)$ group
which contains $D= \frac{n}{2}+2$ dimensional conformal group as its subgroup\footnote{The relation $D=\frac{n}{2}+2$ between the space-time dimension $D$ and $n$ is valid only for $n=2,4,8$ and 16. This is related to the fact that in these dimensions the massless momentum condition $P_mP^m=0$ has the general twistor-like solution $P^m=\bar\lambda\gamma^m\lambda$, where commuting spinors $\lambda_\alpha$ are transformed under a fundamental representation of the D-dimensional conformal group whose subgroup is $Sp(n)$.}. The parameters $a^{\mu \nu}$,
$l_{\mu}{}^{\nu}$ and $k_{\mu \nu}$ contain conventional
translations, Lorentz transformations and conformal boosts respectively, whereas the parameter of dilatation is proportional to the trace 
$g_\mu{}^\mu$.

The constant $\Delta$ which is present in the equation
\p{sp8fbD} is a generalized conformal weight or the $Sp(2n)$ weight of the scalar hyperfield. It is related to the conventional conformal weight of the scalar fields in the corresponding space-time dimensions $D=\frac n2 +2$ as follows \cite{Florakis:2014aaa}
\be\label{confd}
\Delta_D=\frac n2\Delta.
\ee
This is because the D-dimensional dilatation parameter $g_D$ is  $g_D=\frac 2n g_\mu{}^\mu$ 

The free field $b(X)$ satisfies the field equations
\begin{equation}\label{BF}
(\partial_{\mu \nu} \partial_{ \rho \lambda}- \partial_{\mu \rho} \partial_{\nu \lambda}) b(X)=0.
\end{equation}
These equations are invariant under the $Sp(2n)$ transformations \p{sp8fbD}
provided that $\Delta = \frac{1}{2}$, which is therefore a canonical dimension of $b(X)$ \cite{Vasiliev:2001dc}.

\subsection{Spinor field $f_\alpha (X)$}
The half-integer spin fields are packed in a Grassmann-odd ``spinor" hyperfield $f_\alpha (X)$ transforming under the linear representation of $GL(n)$. Under $Sp(2n)$ it transforms as follows
\begin{eqnarray}\label{sp8fer}
\delta f_\alpha(X)&=&-{\Large(}a^{\mu \nu} \partial_{\mu \nu} + \Delta\,(g_\mu{}^\mu-
 k_{\mu \nu}  X^{\mu \nu}) + 2 g_\nu{}^\mu X^{\nu \rho}\,
\partial_{\mu \rho} -
 k_{\mu \nu}  X^{\mu \rho} X^{\nu \lambda}\partial_{\rho \lambda}{\large)} f_\alpha(X)\nonumber\\
 &-&(g_{\alpha}^\beta-k_{\nu\alpha }  X^{\nu \beta})f_\beta\,.
\end{eqnarray}
$f_\alpha(X)$ satisfies the free equations of motion
\begin{equation}\label{feom}
\partial_{\mu[\nu}f_{\alpha ]}=0,
\end{equation}
where $[\alpha\beta]$ denotes anti-symmetrization of indices, while $(\alpha\beta)$ will indicate symmetrization.

These equations are $Sp(2n)$ invariant provided that the field $f_\alpha(X)$ has the generalized conformal dimension
\begin{equation}\label{Deltaf}
\Delta_{\frac 12}=\Delta+\frac 1n=\frac 12+\frac 1n \qquad\Longrightarrow \qquad\Delta=\frac 12\,.
\end{equation}

\subsection{$O(N)$ current $J_{\alpha \beta}(X)$}

Let us assume that the fields $b^A(x)$ and $f^A_\alpha(X)$ belong to a vector representation of the $O({N})$ group ($A=1,\ldots, N)$ and define a generalised current
\begin{align}\label{C-S}\begin{aligned}
J^{AB}_{\alpha \beta}&= b^A \partial_{\alpha \beta} b^B -  b^B \partial_{\alpha \beta} b^A\,,\\
J^{AB}_{\alpha \beta}&=f_\alpha^Af_\beta^B+f_\beta^Af_\alpha^B\,.
\end{aligned}
\end{align}
The current is symmetric with respect to its indices $(\alpha,\beta)$ and anti-symmetric in $A,B$.
Using \eqref{deltaX}, (\ref{sp8fbD}) and \eqref{sp8fer}, one can show that the $Sp(2n)$ transformations of the current are
\begin{align}\label{TRC1}
\delta_a J_{\alpha \beta}^{AB}&=- a^{\mu \nu} \partial_{\mu \nu} J^{AB}_{\alpha \beta}
\\ \label{TRC2}
\delta_g  J_{\alpha \beta}^{AB}& =- \left ( \frac{n+2}{n} \,g_\mu{}^\mu + 2 g_\nu{}^\mu X^{\nu \rho}
\partial_{\mu \rho} \right )
J_{\alpha \beta}^{AB} -l_\alpha{}^\mu J_{\mu \beta}^{AB}
 -  l_\beta{}^\mu J_{\alpha \mu}^{AB}
\\ \label{TRC3}
\delta_k  J_{\alpha \beta}^{AB} &=
( k_{\mu \nu}  X^{\mu \nu}+
 k_{\mu \nu}  X^{\mu \rho} X^{\nu \lambda}\partial_{\rho \lambda})
J_{\alpha \beta}^{AB}
+ k_{\alpha \mu} X^{\mu \nu} J_{\nu \beta}^{AB}+
 k_{\beta \mu} X^{\mu \nu} J_{\alpha \nu}^{AB}
\end{align}
As it can be seen form \p{TRC2} (and \p{TRC3}),
the canonical conformal weight of the current is
\begin{equation}\label{Jd}
\Delta_1=1+\frac 2n\,,
\end{equation}
where the subscript of $\Delta_1$ labels the ``spin" $s=1$ of $J_{\alpha\beta}$ (see also \eqref{Deltaf}). In general, the ``spin" of a symmetric tensor $T_{\alpha_1\ldots\alpha_r}$ of rank $r$ is defined as $s=\frac r2$. This is a natural extension of the notion of spin of fields in $D=3,4$ described by spin-tensors.

One can show that the  current (\ref{C-S}) satisfies the generalized conservation
conditions introduced in \cite{Vasiliev:2002fs}
\begin{equation}\label{conservation-j}
\partial_{\mu \nu}J^{AB}_{\alpha \beta }-\partial_{\mu \alpha}J^{AB}_{\nu \beta } -
\partial_{\beta \nu}J^{AB}_{\alpha \mu } +
\partial_{ \alpha \beta}J^{AB}_{\mu \nu }=0 \qquad \Longleftrightarrow \qquad\YoungpBB=0
\end{equation}
provided that the fields $b(X)$ and $f_\alpha(X)$ satisfy the free field equations (\ref{BF}) and \eqref{sp8fer}, respectively.
On the right in \p{conservation-j}
we indicated the Young symmetry in the indices $(\mu\nu\alpha\beta)$ of the left hand side which is annihilated by the conservation condition.

Note that, in the general case, even if the current $J_{\alpha \beta}$ is not composed of the matter fields, but satisfies the conservation law \eqref{conservation-j}, the  $Sp(2n)$ invariance of the latter requires that its 
generalized conformal dimension is always canonical \eqref{Jd}, which is in accord with the usual CFTs.

\subsection{Stress tensor
$T_{\alpha \beta \gamma \delta}(X)$}

Similarly one can define a free generalised stress (or energy-momentum)
tensor \cite{Vasiliev:2002fs}
\begin{equation} \label{EMT-S}
T_{\alpha \beta \gamma \delta} = \tilde T_{(\alpha \beta, \gamma \delta)},
\end{equation}
with
$$
 \tilde T_{\alpha \beta, \gamma \delta}= (\partial_{\alpha \beta} b)  (\partial_{\gamma \delta} b)-
\frac{1}{3} b \partial_{\alpha \beta} \partial_{\gamma \delta} b\,
$$
or for the fermionic field $f_\alpha$
$$
\tilde T_{\alpha \beta, \gamma \delta}=f_\gamma\partial_{\alpha\beta}f_\delta.
$$

One can check that the totally symmetric tensor
$T_{\alpha \beta \gamma \delta}$ transforms covariantly
under the $Sp(2n)$ transformations (\ref{sp8fbD}) as follows
\begin{align} \label{TRT1}
\delta_a T_{\alpha \beta \gamma \delta}&=- a^{\mu \nu} \partial_{\mu \nu} T_{\alpha \beta \gamma \delta},
\\ \label{TRT2}
\delta_g T_{\alpha \beta \gamma \delta}&=- \left ( \frac{n+4}{n}\,g_\mu{}^\mu + 2 g_\nu{}^\mu X^{\nu \rho}
\partial_{\mu \rho} \right )
T_{\alpha \beta \gamma \delta} -4l_{(\alpha}{}^\mu T_{\beta \gamma \delta)\mu }\,,
\\ \label{TRT3}
\delta_k T_{\alpha \beta \gamma \delta}&=( k_{\mu \nu}  X^{\mu \nu}+
 k_{\mu \nu}  X^{\mu \rho} X^{\nu \lambda}\partial_{\rho \lambda})
T_{\alpha \beta \gamma \delta}
+ 4 k_{\mu(\alpha } X^{\mu \nu} T_{ \beta \gamma \delta)\nu}.
\end{align}
 From the form of the above transformations we see that
the canonical conformal dimension of $T_{\alpha \beta, \gamma \delta}$ is
\begin{equation}\label{Td}
\Delta_2 =1+\frac 4n\,.
\end{equation}
One can also check that the stress tensor (\ref{EMT-S}) satisfies generalized conservation
conditions \cite{Vasiliev:2002fs}
\begin{equation}\label{conservation}
\partial_{\mu \nu}T_{\alpha \beta \gamma \delta}-\partial_{\mu \alpha}T_{\nu \beta \gamma \delta} -
\partial_{\beta \nu}T_{\alpha \mu \gamma \delta} +
\partial_{ \alpha \beta}T_{\mu \nu \gamma \delta}=0 \qquad \Longleftrightarrow \qquad \YoungpDB=0
\end{equation}
provided that the fields $b(X)$ and $f_\alpha(X)$ satisfy the free field equations (\ref{BF}) and \eqref{sp8fer}, respectively. Again, we indicated on the right hand side of \p{conservation}
the relevant symmetry  of $\pl T$ that is set to zero by the $Sp(2n)$ conservation condition.

As in the case of the conserved current $J_{\alpha \beta}(X)$, the $Sp(2n)$ invariance of the conservation law \eqref{conservation} always requires that the conformal dimension of $T_{\alpha \beta \mu \nu}(X)$ is canonical \eqref{Td}.

By analogy with $J_{\alpha\beta}$ and $T_{\alpha\beta\gamma\delta}$ one can introduce higher-spin conserved currents $T_{\alpha_1\ldots \alpha_{2s}}$ ($2s=1,2,3,\ldots$) \cite{Vasiliev:2002fs} which transform under $Sp(2n)$ as follows
\begin{align} \label{TRT1s}
\delta_a T_{\alpha_1 \ldots \alpha_{2s}}&=- a^{\mu \nu} \partial_{\mu \nu} T_{\alpha_1 \ldots \alpha_{2s}},
\\ \label{TRT2s}
\delta_g T_{\alpha_1 \ldots \alpha_{2s}}&=- ( \Delta_s\,g_\mu{}^\mu + 2 g_\nu{}^\mu X^{\nu \rho}
\partial_{\mu \rho})T_{\alpha_1 \ldots \alpha_{2s}} -2sg_{(\alpha_1}{}^\mu T_{\alpha_2 \ldots \alpha_{2s})\mu }\,,
\\ \label{TRT3s}
\delta_k T_{\alpha_1 \ldots \alpha_{2s}}&=( k_{\mu \nu}  X^{\mu \nu}+
 k_{\mu \nu}  X^{\mu \rho} X^{\nu \lambda}\partial_{\rho \lambda})
T_{\alpha_1 \ldots \alpha_{2s}}
+ 4 k_{\mu(\alpha_1 } X^{\mu \nu} T_{\alpha_2 \ldots \alpha_{2s})\nu},
\end{align}
where
\be\label{Ds}
\Delta_s=1+\frac {2s}n\,.
\ee
The $Sp(2n)$ conservation condition \cite{Vasiliev:2002fs} sets to zero the most anti-symmetric component of $\pl T$:
\begin{equation}\label{conservation-s}
\begin{aligned}&\partial_{\mu \nu}T_{\alpha \beta \gamma(2s-2)}-\partial_{\mu \alpha}T_{\nu \beta \gamma(2s-2)} -\\
&\partial_{\beta \nu}T_{\alpha \mu \gamma(2s-2)} +
\partial_{ \alpha \beta}T_{\mu \nu \gamma(2s-2)}=0
\end{aligned}\qquad \Longleftrightarrow \qquad \parbox{60pt}{\bep(60,20)\put(0,0){\YoungBB}\put(20,10){\RectARow{4}{$2s-2$}}\eep}=0
\end{equation}
We will see that when applied to correlation functions these conservation conditions restrict the structure of the former to those of free theories.  Heuristically, 
this happens because the $Sp(2n)$ conservation leads to an over-determined system of equations, i.e. the number of equations  generally exceeds  the number of the components of $T$, as it can be seen from the Young diagram above. On the  contrary, in the usual CFTs the conservation condition
\begin{align}
\pl^m J_{m a_2...a_s}=0 \qquad \Longleftrightarrow\qquad \parbox{40pt}{\RectARow{4}{$s-1$}}=0
\end{align}
has fewer  components than the current $J_{m_1...m_s}$. For $s=1,2$ the conservation is not too restrictive, while for $s>2$ it requires an advanced machinery of Ward identities \cite{Maldacena:2011jn} or higher-spin algebras \cite{Boulanger:2013zza}, see also \cite{Alba:2013yda,Stanev:2013qra,Alba:2015upa}, to see that the presence of at least one higher-spin current implies the presence of full infinite-dimensional higher-spin symmetry of the model and makes it free.

Now we are in a position to consider various $Sp(2n)$
invariant two-, three-, and four-point functions. As we mentioned in the Introduction, when computing the correlation functions we will, a priori, assume
that the fields $b(x)$ and $f_\alpha(X)$
may have arbitrary $Sp(2n)$ weights $\Delta_0=\Delta$ and $\Delta_{\frac 12}=\Delta+\frac 1n$, respectively, with anomalous (spin-independent) dimensions $\Delta$
 and then see how the $Sp(2n)$ invariance and conservation laws restrict their values.

\section{$Sp(2n)$ invariance of multi-point correlation functions}\label{mpcf}
Consider a generic correlation function of $k$  rank-$r_i$ tensor fields $\Phi^{\Delta^{(i)}}(X_i)$ $(i=1,\ldots,k)$ whose spin-independent parts  of the conformal weights are $\Delta^{(i)}$
\begin{equation}\label{multipoint}
\langle\Phi^{\Delta^{(1)}}_{\alpha_1\ldots \alpha_{r_1}}(X_1) \ldots \Phi^{\Delta^{(k)}}_{\beta_1\ldots \beta_{r_k}}(X_k)\rangle \equiv G_{\alpha_1\ldots \alpha_{r_1},\ldots,\beta_1 \ldots\beta_{r_k}}(X_1, \ldots,X_k)\,.
\end{equation}
The correlation function is  invariant under the $Sp(2n)$ transformations \eqref{deltaX} if for any values of the $Sp(2n)$ parameters the following equation holds
\begin{eqnarray}\label{deltaG}
&\sum_{i=1}^{k}\left [\Delta_i(g_\mu{}^{\mu}-k_{\mu\nu}X^{\mu\nu}_i)+\delta X^{\mu\nu}_i\frac{\partial}{\partial X^{\mu\nu}_i}\right]G_{\alpha_1\ldots \alpha_{r_1},\ldots,\beta_1 \ldots\beta_{r_k}}(X_1,\ldots,X_k)&\nonumber\\
&+\sum_{j=1}^{_1}(g_{\alpha_j}{}^{\mu_j}-k_{\alpha_{j}\nu}X_1^{\nu\mu_j})\,G_{\mu_1\ldots \mu_j\ldots \mu_{r_1},\ldots,\beta_1\ldots \beta_{r_k}}(X_1, \ldots,X_k)+\cdots &\nonumber\\
&+\sum_{j=1}^{r_k}(g_{\beta_j}{}^{\mu_j}-k_{\beta_{j}\nu}X_k^{\nu\mu_j})\,G_{\alpha_1\ldots \alpha_{r_k},\ldots,\mu_1\ldots \mu_j\ldots \mu_{r_k}}(X_1, \ldots,X_k) =0\,,&
\end{eqnarray}
where the variations $\delta X^{\mu\nu}_i$ are given in \p{deltaX}.
The equation \p{deltaG} plays an important role in the further analysis, since it determines a general condition
for a given multi-point function to be $Sp(2n)$-invariant.

\subsection{Two-point functions} \label{STPF}
\subsubsection{Correlators of two scalars $ \langle b b\rangle $ and two fermions $\langle ff\rangle$}
Let us consider the two-point function of two scalar fields $b^A(X_1)$ and $b^B(X_2)$ carrying $O(N)$ vector indices $A$ and $B$,
 and having weights $\Delta^{(1)}$ and $\Delta^{(2)}$. The condition that this two-point function
   is $Sp(2n)$-invariant under the transformations \p{sp8fbD}  i.e., satisfies \eqref{deltaG}
implies that it is nonzero if
$\Delta^{(1)}=\Delta^{(2)}= \Delta$, and has the following form \cite{Vasiliev:2003jc}
\begin{equation}\label{2p}
  \langle   b^A(X_1) b^B(X_2)    \rangle=  C_{bb}
  (\det |X_{12}|)^{-\Delta} \delta^{AB},
\end{equation}
where $C_{bb}$ is an arbitrary constant,
 $X^{\alpha\beta}_{12}=X^{\alpha\beta}_1-X^{\alpha\beta}_2$ and
\begin{equation}\label{X^-1}
(X^{-1}_{12})_{\alpha\beta}=(X^{\alpha\beta}_1-X^{\alpha\beta}_2)^{-1}\,.
\end{equation}
Analogously, for the two fermionic fields $f^A_\alpha(X_1)$ and $f^B_\beta(X_2)$ one gets \cite{Vasiliev:2003jc}
\begin{equation}\label{2pf}
  \langle   f_\alpha^A(X_1) f_\beta ^B(X_2)    \rangle=  C_{ff}
  (\det |X_{12}|)^{-\Delta} \delta^{AB}(X_{12})^{-1}_{\alpha\beta},
\end{equation}

The correlator of $b(X_1)$ with $f_\alpha(X_2)$ is zero, as are all the multi-point functions with an odd number of $f_\alpha(X)$, since their index structure is not even $GL(n)$ invariant.

For further use let us also give the form of the $Sp(2n)$ variations \eqref{deltaX} of the matrix $(X^{-1}_{12})_{\alpha\beta}$
\begin{eqnarray}\label{X-1}
\delta (X^{-1}_{12})_{\alpha\beta}&=&-(X^{-1}_{12})_{\alpha\gamma}(\delta X_1-\delta X_2)^{\gamma\delta}(X^{-1}_{12})_{\delta\beta}\nonumber\\
&=& -2g_{(\alpha}{}^\gamma (X^{-1}_{12})_{\beta)\gamma}+k_{\gamma(\alpha}(X_1+X_2)^{\gamma\delta} (X^{-1}_{12})_{\beta)\delta}\nonumber\\
&=& -2g_{(\alpha}{}^\gamma (X^{-1}_{12})_{\beta)\gamma}+k_{\alpha\gamma}X_1^{\gamma\delta} (X^{-1}_{12})_{\delta\beta}+ (X^{-1}_{12})_{\alpha\delta}X_2^{\delta\gamma}k_{\gamma\beta}\,.
\end{eqnarray}
Again the variations  $\delta X^{\mu\nu}_1$ and $\delta X^{\mu\nu}_2$ are given in \p{deltaX}.
{}From eq. \eqref{X-1} it follows that the matrix $(X^{-1}_{12})_{\alpha\beta}$, regarded as a bi-local tensor,  is $Sp(2n)$-invariant, i.e.
\begin{eqnarray}\label{X-12}
&&-(\delta X_1^{\mu\nu}\frac{\partial}{X_1^{\mu\nu}}+\delta X_2^{\mu\nu}\frac{\partial}{X_2^{\mu\nu}}) (X^{-1}_{12})_{\alpha\beta}-2g_{(\alpha}{}^\gamma (X^{-1}_{12})_{\beta)\gamma}+ \\ \nonumber
&&+k_{\alpha\gamma}X_1^{\gamma\delta} (X^{-1}_{12})_{\delta\beta}+ (X^{-1}_{12})_{\alpha\delta}X_2^{\delta\gamma}k_{\gamma\beta} =0\, ,
\end{eqnarray}
With the help of relation \p{X-12} one can immediately check $Sp(2n)$ invariance of correlation functions away from the singularity point.
Notice that the conformal boosts act effectively on the first index of $(X^{-1}_{12})_{\alpha\beta}$ with the matrix $X_1$ and on the second index with the matrix $X_2$ (or vice verse). This ensures the same $Sp(2n)$-invariant properties of the left- and right-hand side of \eqref{2pf} in accordance with the generic formula \eqref{deltaG}. The matrix $(X^{-1}_{12})_{\alpha\beta}$, together with its determinant, is thus one of the elementary building blocks of all the $Sp(2n)$-invariant correlation functions.

\subsubsection{Two current correlator $ \langle JJ\rangle$}
The two-point function of the conserved currents $J^{AB}_{\alpha\beta}$
can be derived by writing the most general expression
compatible with its index symmetries, requiring 
its invariance under $O(N)$ and  $Sp(2n)$,  and imposing
the current conservation condition. 
The $Sp(2n)$ invariance condition \eqref{deltaG} reduces a general expression to the one constructed in terms of the matrix $X^{-1}_{12}$ and its determinant as follows
\begin{eqnarray} \label{JJ-2-1-in}
\langle   J^{AB}_{\alpha \beta}(X_1) J^{CD}_{\mu \nu}(X_2)    \rangle
 =C_{JJ} ({\det|X_{12}|})^{-1}  (P_{12})_{\alpha \beta, \mu \nu}(\delta^{AC} \delta^{BD}- \delta^{AD} \delta^{BC})\,,
\end{eqnarray}
where we defined
\begin{equation} \label{defp}
(P_{ab})_{\alpha \beta, \mu \nu}= (X^{-1}_{ab})_{\mu \alpha} (X^{-1}_{ab})_{\nu \beta} + (X^{-1}_{ab})_{\nu \alpha} (X^{-1}_{ab})_{\mu \beta}
\end{equation}
and
\begin{equation}\label{xab}
X^{-1}_{ab} \equiv (X_a^{\alpha \beta} -  X_b^{\alpha \beta})^{-1}\equiv (p_{ab})_{\alpha\beta},
\end{equation}
$a,b=1,2$ and $a\not= b$. 

The constant $C_{JJ}$ is arbitrary, while the power of the determinant is fixed to be $-1$, i.e. its absolute value is equal to 
 the canonical value $\Delta=1$ of the spin independent part of the  conformal weight of  $J_{\alpha \beta}$. One can easily check that \eqref{JJ-2-1-in} obeys the current conservation law \eqref{conservation-j} \footnote{Note that for the currents of the form \eqref{C-S} one can obtain  \p{JJ-2-1-in}  simply using  the two-point function \p{2p} of the free scalars with the canonical dimension $\Delta_0=\frac{1}{2}$ or the two-point function \eqref{2pf} of the two fermions with the canonical dimension $\Delta_{\frac 12}=\frac 12 +\frac 12$.}. 

The $Sp(2n)$ variations of \eqref{defp} are determined by those of $X^{-1}_{12}$, given in \eqref{X-12}, and have the following form
\begin{eqnarray}\label{gp}
&&2 g_\sigma{}^\rho \left (  X_1^{\sigma\lambda}\frac{\partial}{\partial X_1^{\rho \lambda}} + X_2^{\sigma\lambda}\frac{\partial}{\partial X_2^{\rho\lambda}}
\right ) (P_{12})_{\alpha \beta, \mu \nu} = \\ \nonumber
&&=- g_\alpha{}^\sigma (P_{12})_{\sigma \beta, \mu \nu} -  g_\beta{}^\sigma (P_{12})_{\alpha \sigma, \mu \nu}
-g_\mu{}^\sigma (P_{12})_{\alpha\beta, \sigma \nu}
-g_\nu{}^\sigma (P_{12})_{\alpha \beta, \mu \sigma}\,,
\end{eqnarray}
\begin{eqnarray} \label{kp}
&& k_{\rho\sigma} \left (X_1^{\rho\lambda}  X_1^{\sigma\delta}\frac{\partial}{\partial X_1^{\lambda\delta}}  +
X_2^{\rho\lambda}  X_2^{\sigma\delta}\frac{\partial}{\partial X_2^{\lambda\delta}}
\right ) (P_{12})_{\alpha \beta, \mu \nu} = \\ \nonumber
&&=- k_{\alpha \sigma} X_1^{\sigma\delta}(P_{12})_{\delta \beta, \mu \nu} -  k_{\beta \sigma} X_1^{\sigma\delta}(P_{12})_{\alpha \delta, \mu \nu}
- k_{\mu \sigma} X_2^{\sigma\delta}(P_{12})_{\alpha \beta, \delta \nu} -  k_{\nu\sigma} X_2^{\sigma\delta}(P_{12})_{\alpha \beta, \mu \delta}\,.
\end{eqnarray}
It is important for the proof of the $Sp(2n)$ invariance of \eqref{JJ-2-1-in} that in the right hand side of \eqref{kp} the first pair of the indices of $(P_{12})_{\alpha\beta,\gamma\delta}$ gets rotated with the matrix $k_{\alpha\sigma}X^{\sigma\delta}_1$ and the second pair gets rotated with $k_{\mu\sigma}X^{\sigma\delta}_2$ . The  bi-local tensor \eqref{defp} is a building block of correlation functions of higher even-rank tensors such as the stress tensor.

\subsubsection{Stress tensor correlator $  \langle TT \rangle$}
Following the same reasoning as above, one gets the $Sp(2n)$-invariant two-point function of the two stress tensors
\begin{equation} \label{2TTT-1}
\langle   T_{\alpha \beta \gamma \delta}(X_1) T_{\mu \nu \rho \sigma}(X_2)   \rangle=  C_{TT}\frac{1}{\det|X_{12}|}
 \left ( (P_{12})_{\alpha \beta, \mu \nu}(P_{12})_{\gamma \delta, \rho \sigma}+ symm. \right )
\end{equation}
where the total symmetrization of the both sets of indices $(\alpha \beta \gamma \delta)$ and
$(\mu \nu \rho \sigma)$ is assumed. The $Sp(2n)$ invariance \eqref{deltaG} and the conservation properties  of \eqref{2TTT-1} dictated by \eqref{conservation}
can be checked with the use of the form of the variations \eqref{gp}, \eqref{kp} and of $\det |X_{12}|$.

\section{Three-point functions}\label{STHPF}
\subsection{Scalars and fermions} \label{sbbb}
A three-point function for three fields $b(X)$ with weights $\Delta^{(1)}, \Delta^{(2)}$ and $\Delta^{(3)}$ has the following form \cite{Vasiliev:2003jc}
\begin{equation} \label{3pb}
 \langle   b(X_1) b(X_2) b(X_3)    \rangle
= {(\det |X_{12}|)}^{-{k_3}} \,{(\det |X_{23}|)}^{-{k_1}}\,{(\det |X_{13}|)}^{-{k_2}}\,,
\end{equation}
where
\begin{equation} \label{3k-1}
k_1 =\frac{1}{2}( \Delta^{(2)} + \Delta^{(3)} - \Delta^{(1)}), \quad k_2 = \frac{1}{2}(\Delta^{(3)} +\Delta^{(1)} - \Delta^{(2)}), \quad
k_3 = \frac{1}{2}(\Delta^{(1)} + \Delta^{(2)} - \Delta^{(3)})
\end{equation}
Correspondingly, the correlator of two fermions and a scalar is \cite{Vasiliev:2003jc}
\begin{equation} \label{2pfb}
 \langle   f_\alpha(X_1) f_\beta(X_2) b(X_3)    \rangle
= (X^{-1}_{12})_{\alpha\beta}{(\det |X_{12}|)}^{-{k_3}} \,{(\det |X_{23}|)}^{-{k_1}}\,{(\det |X_{13}|)}^{-{k_2}}\,.
\end{equation}
Let us mention that
in a similar way
one can find four-point functions which include four scalars, four fermions and two scalars and two fermions.
As in the ordinary CFTs the $Sp(2n)$ symmetry fixes
the four-point correlators up to an arbitrary function of cross-ratios \cite{Florakis:2014kfa}.

\subsection{ Three-point functions $\langle Jbb\rangle $, $\langle Tbb\rangle $, $\langle Jff\rangle $ and $\langle Tff\rangle $}
One more tensor structure appears in the  $Sp(2n)$--invariant three-point correlators  which involve two scalars of dimensions $\Delta^{(1)}$ and $\Delta^{(2)}$, and one conserved current or stress-tensor of the spin-independent conformal dimension $\Delta^{(3)}=1$
\begin{eqnarray} \label{bbJ-JJ}
&&\langle   b^A(X_1) b^B(X_2) J_{\alpha \beta}^{CD}(X_3)    \rangle = \\ \nonumber
&&=C_{bb J}
{(\det |X_{12}|)^{-k_3}  (\det |X_{13}|)^{-k_2} (\det |X_{23}|)^{-k_1}} (Q^3_{12})_{\alpha \beta}
 (\delta^{AC}\delta^{BD}- \delta^{AD}\delta^{BC})
\end{eqnarray}
and
\begin{eqnarray} \label {bbT-TT} \nonumber
&&\langle   b(X_1) b(X_2) T_{\alpha \beta \gamma \delta}(X_3)    \rangle
=C_{bb T}
{(\det |X_{12}|)^{-k_3}  (\det |X_{13}|)^{-k_2} (\det |X_{23}|)^{-k_1}} \times \\
&& \times ((Q^3_{12})_{\alpha \beta}
(Q^3_{12})_{\gamma \delta}
+ (Q^3_{12})_{\alpha \gamma} (Q^3_{12})_{\beta \delta}
 + (Q^3_{12})_{\alpha \delta}
 (Q^3_{12})_{\beta \gamma}
)
\end{eqnarray}
where constants $k_i$ again satisfy the conditions
\p{3k-1}. In the above expressions
\begin{equation} \label{defq}
(Q^c_{ab})_{\alpha \beta} = (X^{-1}_{ac})_{\alpha \beta} - (X^{-1}_{bc})_{\alpha \beta},
\end{equation}
where $a,b,c=1,2,3$ and $a\not =b\not = c$.

The quantities $(Q^c_{ab})_{\alpha \beta}$ are manifestly invariant under the translation of the coordinates and
transform in the following way under the $GL(n)$ rotations and the generalized boosts
\begin{eqnarray}\label{gq}
&&2 g_\nu{}^\mu \left (  X_1^{\nu \rho}\frac{\partial}{\partial X_1^{\mu \rho}} + X_2^{\nu \rho}\frac{\partial}{\partial X_2^{\mu \rho}} + X_3^{\nu \rho}\frac{\partial}{\partial X_3^{\mu \rho}}
\right ) (Q^3_{12})_{\alpha \beta} = \\ \nonumber
&&=- g_\beta{}^\mu (Q^3_{12})_{\alpha \mu} -  g_\alpha{}^\mu (Q^3_{12})_{\beta \mu}\,,
\end{eqnarray}
\begin{eqnarray}\label{kq}
&& k_{\mu \nu} \left (X_1^{\mu \rho}  X_1^{\nu \tau}\frac{\partial}{\partial X_1^{\rho \tau}}  +
X_2^{\mu \rho}  X_2^{\nu \tau}\frac{\partial}{\partial X_2^{\rho \tau}}  +
X_3^{\mu \rho}  X_3^{\nu \tau}\frac{\partial}{\partial X_3^{\rho \tau}}
\right ) (Q^3_{12})_{\alpha \beta} = \\ \nonumber
&&=- k_{\alpha \rho} X_3^{\rho \tau}(Q^3_{12})_{\tau \beta} -
k_{\beta \rho} X_3^{\rho \tau}(Q^3_{12})_{\tau \alpha}\,.
\end{eqnarray}
For the $Sp(2n)$ invariance \eqref{deltaG} of the correlators \eqref{bbJ-JJ} and \eqref{bbT-TT} it is important to notice that the contraction of the indices of $Q^3_{12}$ on the right hand side of the boost transformations \eqref{kq} only involves the coordinate $X_3$.

Using \eqref{xder} and \eqref{consq1} one can check that \p{bbJ-JJ} and \p{bbT-TT} satisfy the current and stress-tensor conservation laws if
\begin{equation} \label{usl0}
k_1=k_2=\frac{1}{2}
\end{equation}
and $k_3$ is arbitrary. This means that the generalized dimensions of the fields $b(X_1)$ and $b(X_2)$ are equal to each  other but otherwise unrestricted (i.e. can be anomalous).

Correspondingly, for the correlators of two fermions of dimension $\Delta_{\frac 12}=\Delta+\frac 1n$  with the current $J$ and the stress tensor $T$ we have
\begin{eqnarray} \label{ffJ-JJ}
&&\langle  f_\gamma^A(X_1) f_\delta^B(X_2) J_{\alpha \beta}^{CD}(X_3)    \rangle = \\ \nonumber
&&=C_{ff J}
{(\det |X_{12}|)^{\frac {1-\Delta}2}  (\det |X_{13}|)^{-\frac 12} (\det |X_{23}|)^{-\frac 12}} (X_{12})^{-1}_{\gamma\delta} (Q^3_{12})_{\alpha \beta}
 (\delta^{AC}\delta^{BD}- \delta^{AD}\delta^{BC})
\end{eqnarray}
and
\begin{eqnarray} \label {ffT-TT} \nonumber
&&\langle   f_\mu(X_1) f_\nu(X_2) T_{\alpha \beta \gamma \delta}(X_3)    \rangle
=C_{ff T}
{(\det |X_{12}|)^{\frac {1-\Delta}2}  (\det |X_{13}|)^{-\frac 12} (\det |X_{23}|)^{-\frac 12}} \times \\
&& \times (X_{12})^{-1}_{\mu\nu}((Q^3_{12})_{\alpha \beta}
(Q^3_{12})_{\gamma \delta}
+ (Q^3_{12})_{\alpha \gamma} (Q^3_{12})_{\beta \delta}
 + (Q^3_{12})_{\alpha \delta}
 (Q^3_{12})_{\beta \gamma}
)\,.
\end{eqnarray}

Looking at the form of the two- and three-point correlation functions constructed above we come to the conclusion that except for some degenerate cases\footnote{Degenerate cases correspond to the situations in which additional invariants can be built with the help of  $\epsilon$-symbols, as it happens in $3d$ \cite{Giombi:2011rz}.} the most general multi-point function can be written as a sum over all possible polynomials of a required rank in three structures $p_{ab}=X^{-1}_{ab}$  \eqref{xab}, $P_{ab}$ \eqref{defp}\footnote{$P_{ab}$ is actually a square of $p_{ab}=X^{-1}_{ab}$, but it is convenient to regard it as an 
independent structure in order to separate contributions from bosons and fermions (see Section \ref{HSalg} for more details).} and $Q^c_{ab}$ \eqref{defq}  times a pre-factor which in the case of four-point and higher order
correlators  is a function of $Sp(2n)$-invariant cross-ratios:
\begin{align}
\langle \Phi...\Phi\rangle&= G(p_{ab},P_{ab},Q^c_{ab}|X_{ab})
\end{align}
This statement is completely analogous to the one for the usual CFTs (see \cite{Costa:2011mg} for a proof). We will discuss the general structure of the three-point correlators in more detail in Section \ref{HSalg}.

\section{Three-point functions with fixed conformal dimension} \label{SOTPF}
We will now consider $Sp(2n)$-invariant three-point functions in which the requirement of the current and stress-tensor conservation completely fixes the conformal dimension of the scalar operator to be 1, i.e. twice that of the canonical dimension of the free scalar hyperfield. At this point the 
 properties of correlators in
$Sp(2n)$--invariant systems (for $n>2$) become different from those of conventional $D$-dimensional CFTs where in the analogous correlators a restriction on the dimension of the scalar operator does not occur. The obtained restriction  suggests that the $Sp(2n)$-invariant systems under consideration are free, as we will discuss in more detail below.

\subsection{$\langle JJ\mathcal O\rangle$ three-point functions}
The simplest three-point function of this kind is $\langle JJ\mathcal O \rangle$, where $\mathcal O(X)$ is a scalar operator of dimension $\Delta$ which, in general, can be a composite of the elementary fields $b(X)$. From the requirement
of $Sp(2n)$ invariance one finds that the correlator has the following form
\begin{align} \label{jjo}
 \langle   J_{\mu \nu }(X_1) \mathcal O(X_2) J_{\alpha \beta}(X_3)    \rangle =&{(\det |X_{12}|)^{-\frac \Delta 2}  (\det |X_{13}|)^{-\frac{2-\Delta}2} (\det |X_{23}|)^{-\frac\Delta 2}} \times \\
& \times \left( {\cal A} [ (Q^3_{12})_{\alpha \beta} (Q^1_{23})_{\mu\nu}]
+ {\cal B}(P_{13})_{ \mu\nu,\alpha\beta}\right) \nonumber
\end{align}
where ${\cal A}$ and $\cal B$
are some yet undetermined constants.

Now let us impose the current conservation condition \p{conservation-j}
on the three-point function \p{jjo}.
Requiring, for example, the conservation of the current $J_{\mu \nu}(X_1)$
one gets (see the Appendix
\ref{Appendix C}
for details)
\begin{equation} \label{usl1}
{\cal A}={\cal B}, \,\,\,\,\,\,  \Delta=1
\end{equation}
and similarly for the conservation of the current
$J_{\alpha \beta}(X_3)$.
We thus conclude that for the three-point function \eqref{jjo} to be non-zero the dimension of the scalar operator $\mathcal O$  must be equal to one. 

 There at least two interpretations of this result.
and we will soon find more examples of the same kind. 
Firstly, one can start from a free theory \eqref{BF} with a fundamental field $b(X)$ and see if any interactions are possible. Here we do not assume that interactions admit any realization in terms of either Lagrangian or equations of motion. Then the above constraint implies that $\langle JJ[b^k]\rangle=0$ for $k=1,3,4,...$ in the interacting theory, where $[b^k]$ is a quasi-primary field built of $k$ fields $b(X)$. The only non-zero correlator corresponds to $k=2$ and does not allow for any anomalous dimension. The property of allowing for anomalous dimensions is the most important property of any CFT. Therefore, we interpret our result as the fact that there are no nontrivial interactions possible for the field $b(X)$. Secondly, one can start with any $Sp(2n)$ invariant CFT and assuming that it has a scalar operator ${\mathcal O}(X)$ and a conserved current $J_1(X)$, one concludes 
that there are no anomalous dimensions possible for the operator ${\mathcal O} (X)$ as well as for the operators
$[{\mathcal O}^k(X)]$. Moreover, if the correlator $\langle JJ {\mathcal O}\rangle$ is different from zero then the correlator looks like the one for ${\mathcal O}(X)=b^2(X)$, where $b(X)$ is a free field.

This situation is different from the one in $D \geq 3$  CFTs (see e.g. \cite{Costa:2011mg,Osborn:1993cr,Erdmenger:1996yc}) where an analogous three-point function does not impose any condition on the dimensions of the scalar fields. Using the same ansatz as above with the understanding that each pair of the  indices $(\alpha\beta)$ should be replaced by one vector index of the Lorentz group $SO(1,D-1)$ one gets the condition
\begin{align}
{\cal A} (D-1-\Delta)-{\cal B}\Delta=0
\end{align}
which relates the three parameters but does not impose any restriction on the conformal weight $\Delta$.

The reason  for this difference is that in the $Sp(2n)$-invariant systems in the hyperspace with {\it extra} coordinates the conservation conditions are more restrictive than in
ordinary CFTs. The situation does not change even if we consider weaker conservation conditions, e.g. only an $Sp(n)$-invariant part of \eqref{conservation-j} and \eqref{conservation} obtained by contracting the latter with the symplectic metrics $C^{\alpha\nu}C^{\beta\mu}$.
The exception is the case of $n=2$, $D=3$ in which the $Sp(4)\sim SO(2,3)$
symmetry simply coincides with the three-dimensional conformal symmetry. In this case
the condition \p{usl1} does not arise due to extra $2\times 2$ matrix identities
which are present in $D=3$ as it is shown in  Appendix \ref{Appendix C}.

We can also consider a correlator which contains  two currents and an antisymmetric tensor operator $\mathcal O_{[\alpha\beta]}$.
This operator  can be constructed e.g. by taking the product of two fermionic fields $f_\alpha (X_2) f_\beta (X_2)$. Such a correlator has the following structure
\begin{align}
  \langle   J_{\alpha_1\alpha_2}(X_1)
 & J_{\beta_1\beta_2}(X_2) \mathcal O_{[\gamma_1\gamma_2]}(X_3)  \rangle ={ (\det |X_{12}|)^{-\frac{2-\Delta}2} (\det |X_{13}|)^{-\frac \Delta 2} (\det |X_{23}|)^{-\frac\Delta 2}} \times &\\ \nonumber
& \times  \mathcal C\,\Big [(X^{-1}_{12})_{\beta_1(\alpha_1}(X^{-1}_{13})_{\alpha_2)[\gamma_1}(X^{-1}_{23})_{\gamma_2]\beta_2}+(X^{-1}_{12})_{\beta_2(\alpha_1}(X^{-1}_{13})_{\alpha_2)[\gamma_1}(X^{-1}_{23})_{\gamma_2]\beta_1}\Big].&
\end{align}
Again the  requirement of current conservation  fixes the conformal dimension of $\mathcal O_{[\gamma_1\gamma_2]}$ to be $\Delta=1$.

\subsection{$\langle TJ\mathcal O\rangle $ and $\langle TT\mathcal O\rangle $ correlators}
Other examples of correlation functions in which the conformal weights of the operators are completely fixed by the conservation laws are $\langle TJ\mathcal O \rangle$,  and $\langle TT\mathcal O\rangle$.

The $Sp(2n)$ invariance restricts the correlation function
$\langle TJ\mathcal O \rangle$ to have the following form
\begin{align}
  \langle   T_{\alpha(4) }(X_1) J_{\beta (2)}(X_2)\mathcal O_{\Delta}(X_3)    \rangle =&{(\det |X_{12}|)^{-\frac{2-\Delta}2}  (\det |X_{13}|)^{-\frac\Delta 2} (\det |X_{23}|)^{-\frac\Delta 2}} \\ \nonumber
&\times \left( {\cal A} [ (Q^1_{23})_{\alpha \alpha} (Q^1_{23})_{\alpha\alpha}(Q^2_{13})_{\beta\beta}]
+ {\cal B}(Q^1_{23})_{\alpha\alpha}(P_{12})_{ \alpha\alpha,\beta\beta}\right)\,
\end{align}
where the total symmetrization of the  indices denoted by the same letter is implied.
The conservation  of the stress-tensor $\pl_1 (T_{\alpha(4) })$ and the current $\pl_2 (J_{\beta(2) })$ require
\begin{equation}
{\cal A}=-{\cal B}/2\,, \qquad \Delta=1\,.
\end{equation}
 Therefore we have found again  that the structure of the three-point function and the conformal dimension of the scalar field are fixed by the conservation laws.

The same happens with the $Sp(2n)$-invariant correlator
$\langle TT\mathcal O \rangle$ which has the  form
\begin{eqnarray}
 \langle   T_{\alpha(4) }(X_1) T_{\beta(4)}(X_2) \mathcal O_{\Delta} (X_3)    \rangle  &={(\det |X_{12}|)^{-\frac{2-\Delta}2}  (\det |X_{13}|)^{-\frac\Delta 2} (\det |X_{23}|)^{-\frac\Delta 2}}&\nonumber\\
& \times\Big [ {\cal A} \,(Q^1_{23})_{\alpha \alpha} (Q^1_{23})_{\alpha\alpha}(Q^2_{13})_{\beta\beta}(Q^2_{13})_{\beta\beta}+&\nonumber\\
&+ 
{\cal B}\,(Q^1_{23})_{\alpha\alpha}(Q^2_{13})_{\beta\beta}(P_{12})_{ \alpha\alpha,\beta\beta}]&\nonumber\\
&+{\cal C}\,(P_{12})_{ \alpha\alpha,\beta\beta}(P_{12})_{ \alpha\alpha,\beta\beta}\Big ].& 
\end{eqnarray}
And the conservation of the stress
tensor  fixes the parameters as follows
\begin{equation}
{\cal A}=-{\cal B}/4\,, \qquad {\cal C}=-{\cal B}/6\,, \qquad \Delta=1\,.
\end{equation}

\section{Generic structure of the three-point correlation functions of symmetric tensor operators} \label{HSalg}
The form of the $Sp(2n)$-invariant correlator of three currents $J^{i}_{\alpha\beta}$, where $i$ is an internal group index (e.g. $i$ stands for $AB$ in the $O(N)$ case) manifests a general structure of the three-point functions which include three tensor structures which we called 
$p_{ab}$, $P_{ab}$ and $Q^c_{ab}$ (given in \p{xab}, \p{defp} and \p{defq}, respectively)
\begin{eqnarray}\label{bfJJJ}
&&  \langle   J^{i}_{\alpha_1\alpha_2}(X_1)
  J^j_{\beta_1\beta_2}(X_2) J^k_{\gamma_1\gamma_2}(X_3)    \rangle =\frac{f^{ijk}}{(\det |X_{12}|)^{k_3}  (\det |X_{13}|)^{k_2} (\det |X_{23}|)^{k_1}} \times \\ \nonumber
&& \times  \Big [ {\cal A}(Q^1_{23})_{\alpha_1 \alpha_2} (Q^2_{13})_{\beta_1\beta_2}(Q^3_{12})_{\gamma_1\gamma_2}
+ {\cal B}\Big((Q^3_{12})_{\gamma_1\gamma_2}(P_{12})_{ \alpha_1\alpha_2,\beta_1\beta_2} +  
\text{perm.~of~1,2,3}\Big) \nonumber\\
&&+\mathcal C\,\Big ((p_{12})_{\beta_1(\alpha_1}(p_{13})_{\alpha_2)(\gamma_1}(p_{23})_{\gamma_2)\beta_2}+
(p_{12})_{\beta_2(\alpha_1}(p_{13})_{\alpha_2)(\gamma_1}(p_{23})_{\gamma_2)\beta_1})\Big)\Big],\nonumber
\end{eqnarray}
where $f^{ikj}$ are structure constants of the internal group.
As in the conventional CFTs, the first two structures entering \eqref{bfJJJ} with the parameters $\mathcal A$ and $\mathcal B$ are associated with the currents constructed with the bosonic fields,  while the third structure is associated with the form of the fermionic currents \eqref{C-S}.

The current conservation leaves the parameter $\mathcal C$ arbitrary. It fixes the parameters $k_a$ and relates the parameters of the bosonic structures
\begin{align}  
k_1=k_2=k_3=1/2, \qquad {\cal A}=-{\cal B},\,
\end{align}
which is consistent with the fact that the canonical spin-independent part of conformal dimension of the current is $\Delta=1$. The above correlator has the same structure as in the ordinary CFTs, expect for the $3D$ case where an extra odd structure exists \cite{Giombi:2011rz}.

We are now in a position to discuss the general structure of the three-point correlators of conserved currents which are symmetric tensors of a rank $r=2s$ with $s$ being an integer `spin'. To this end it is convenient to hide the tensor indices away by contracting them with auxiliary variables $\lambda^\ga_a$, where $a$ refers to the point of operator insertion:
\begin{align}
(p_{ab})_{\alpha\beta}&\Rightarrow p_{ab}=(X^{-1}_{ab})_{\alpha\beta}\, \lambda^\alpha_a \lambda_b^\beta \quad \text{no summation over }a,b\,.\\
(P_{bc})_{\alpha\beta,\gamma\delta}&\Rightarrow P_{ab}= 2 p_{ab}p_{ba}=(P_{ab})_{\alpha\beta,\gamma\delta}\, \lambda^\alpha_a \lambda^\beta_a \lambda^\gamma_b \lambda^\delta_b \quad \text{no summation over }a,b\,,\\
(Q^{a}_{bc})_{\alpha\beta}&\Rightarrow Q^a_{bc}= (Q^{a}_{bc})_{\alpha\beta}\, \lambda^\alpha_a \lambda^\beta_a\quad \text{no summation over }a\,.
\end{align}
For instance the correlator of two scalar operators $\mathcal O$ of the same dimension $\Delta$ with a conserved current of an integer spin $s$ obeying \eqref{conservation-s} is
\begin{eqnarray} \label{bbJ-HS2}
\langle   O(X_1) O(X_2) J_{s}(X_3)    \rangle =C
{(\det |X_{12}|)^{-\frac{2-\Delta}2}  (\det |X_{13}|)^{-\frac 12} (\det |X_{23}|)^{-\frac 12}} (Q^3_{12})^s
\end{eqnarray}
Imposing the current conservation condition leads to the same result as for the currents with $s=1,2$, i.e. $k_1=k_2=\tfrac12$, which 
means that the dimensions of the scalar operators are arbitrary.

However, if we consider a three-point function of a scalar operator and two conserved currents
$$
J_s=J_{\alpha_1\ldots\alpha_{2s}}\lambda^{\alpha_1}\cdots\lambda^{\alpha_{2s}}
$$
of ranks $2s_1$ and $2s_2$ with $s\geq 1$ we will again find that, up to an  overall factor, all the free parameters in the correlator are fixed. For example, 
\begin{equation} \label{bbJ-HS3}
\langle   J_3(X_1) J_1(X_2) O(X_3)    \rangle =C\frac{{(Q^1_{23})^3 Q^2_{13}-3 (Q^1_{23})^2 P_{12} }}{\Big(\det |X_{12}|\det |X_{13}|\det |X_{23}|\Big)^{1/2}} 
\end{equation}
is the unique solution of the $Sp(2n)$ conservation conditions \eqref{conservation-s}.

Summarizing, we have found that the restrictions imposed by the $Sp(2n)$ conservation laws on the correlation functions fix their structure up to an overall factor and the  correlators are those of free $Sp(2n)$ invariant CFT. We studied thoroughly the correlators which include   generalised currents and stress-tensor. We can make further progress by computing correlators involving higher-rank tensors. However the study of  examples with higher-spin currents 
 do not lead to any new conclusions comparing to the
  the study of the correlation functions involving a rigid symmetry current $J_1$ and the stress tensor $T_2$.
 In particular, the correlators with higher-spin currents also look like those 
 in a free $Sp(2n)$ invariant CFT. This suggests that all other correlators with higher-spin currents follow the same pattern. The generating function of correlators in free theories were already obtained  \cite{Giombi:2010vg,Colombo:2012jx,Didenko:2012tv,Gelfond:2013xt,Didenko:2013bj}. For example, a generating function of the three-point functions of currents built out of free scalars $b(X)$ is
\begin{equation}
    \langle JJJ\rangle=\frac{\cos(p_{12})\cos(p_{13})\cos(p_{23})\,\exp\left(\frac12[Q^1_{23}+Q^2_{13}+Q^3_{12}]\right)}{(\det |X_{12}|\det |X_{23}|\det |X_{13}|)^{1/2}}\,.
    \end{equation}
It contains operators $J_{s}$, $s=0,1,2,...$ and the correlator $\langle J_{s_1} J_{s_2} J_{s_3}\rangle$ is obtained as the coefficient in front of $(\lambda_1)^{2s_1}(\lambda_2)^{2s_2}(\lambda_3)^{2s_3}$. 

The generating function obtained from the currents built out of free fermions $f_\alpha(X)$ is
\begin{equation}
    \langle JJJ\rangle=\frac{\sin(p_{12})\sin(p_{13})\sin(p_{23})\,\exp\left(\frac12[Q^1_{23}+Q^2_{13}+Q^3_{12}]\right)}{(\det |X_{12}|\det |X_{23}|\det |X_{13}|)^{1/2}}\,.
\end{equation}
The generating function of multi-point correlators can be found in \cite{Didenko:2012tv,Gelfond:2013xt,Didenko:2013bj}. 

The above expressions deal with the bosonic symmetric tensor currents of even rank. The generating function which produces 3-point correlators involving two fermionic currents of odd ranks is similar, see e.g. \cite{Maldacena:2011jn}.

\section{Conclusion} \label{Conclusions}
We have studied restrictions imposed by the generalized conformal group $Sp(2n)$ and the  conservation laws on various correlation functions involving conserved currents, stress-tensor and higher-spin currents. The general structure of the correlators is similar to the one in the usual conformal field theories. It is build of three conformally-invariant tensor structures, which were found by working out the simplest two- and three-point functions. 

The difference between $Sp(2n)$- and $SO(2,D)$-invariant correlation functions arises when conserved tensor currents are involved. If we assume that the theory has a conserved stress-tensor $T$ and, possibly, a conserved current $J$, then computation shows that while the simplest correlators $\langle T\mathcal{OO}\rangle$ and $\langle J\mathcal{OO}\rangle$ allow operators $\mathcal O$ to have an anomalous dimension, their correlators with two conserved tensors, $\langle TTO\rangle$, $\langle JJ\mathcal O\rangle$ and $\langle TJ\mathcal O\rangle$, fix the conformal dimension of $\mathcal O$ to be twice that of a free field.  This is not the case in the $SO(2,D)$ CFTs in which similar restrictions arise only if they contain in addition to the stress tensor a conserved higher-spin current \cite{Maldacena:2012sf,Boulanger:2013zza,Alba:2013yda,Stanev:2013qra,Alba:2015upa}.

The fact that the above restriction on the conformal weight of $\mathcal O$ implies that the theory is free can also be understood as follows. If a free theory contains, e.g. elementary scalar fields $b(X)$ of the canonical dimension 1/2 then we can take $\mathcal O=bb$. In the $Sp(2n)$--invariant free theory we expect that  $\langle TTO\rangle\not =0$, which is already true for a free field. If a free $Sp(2n)$--invariant model could be deformed by some interactions with a coupling constant $g$, quantum corrections would make fields to acquire anomalous dimensions $\Delta(g)$. If so, then as we have seen, the conservation of $T$ requires that the correlator  $\langle TTO_{\Delta(g)}\rangle=0$. Such a theory would not have a smooth free limit $g\rightarrow 0$, since at $g=0$ the correlator is non-zero. 

Therefore, for $n>2$ the rigid $Sp(2n)$ conservation condition turns out to be much more restrictive than the $SO(2,D)$ one. Only in the $n=2$, $D=3$ case the $Sp(4)\sim SO(2,3)$ becomes the usual three-dimensional conformal symmetry and the conservation condition reduces to the usual current conservation which does not restrict conformal dimension of operators in the correlators with spin-one and spin-two currents.

Assuming the presence in the $Sp(2n)$-invariant theory, containing a stress-tensor, of at least one higher-spin current one can easily repeat the proof given in \cite{Maldacena:2012sf,Boulanger:2013zza,Alba:2013yda,Stanev:2013qra,Alba:2015upa} and conclude that there should be infinitely many (symmetric) higher-spin currents associated with a unique higher-spin algebra that is generated by the corresponding charges. In this sense our work shows that in the $Sp(2n)$ setup with $n>2$  spin-one and spin-two currents already behave like higher-spin currents, forcing the theory to be a free one. If we do not assume the existence of the conserved stress-tensor, like in gravity theories, then the above reasoning does not apply. However, the possibility of introducing `hypergravity' interactions of the $Sp(2n)$-invariant systems is an open problem itself. 

Another option that may still lead to interacting $Sp(2n)$-theories without contradicting our results is the existence of certain contact terms in the $Sp(2n)$ correlators, i.e. the terms that have $\delta$-like singularity when points collide.  

Therefore, the main conclusion is that the generalized conformal field theories with
$Sp(2n)$ symmetry only admit a free field realization, with the few loopholes mentioned above. Still the $Sp(2n)$--invariant formulation can be  useful for the study of free CFTs. In particular, one can derive the generating functions for all the correlation functions \cite{Giombi:2010vg,Colombo:2012jx,Didenko:2012tv,Gelfond:2013xt,Didenko:2013bj} and work out the operator algebra \cite{Gelfond:2013xt}. In order to allow for nontrivial interactions the $Sp(2n)$ symmetry should be broken, as  happens for the current interactions \cite{Gelfond:2015poa}.

In this paper we have mainly studied the correlation functions of scalar operators with conserved currents that are totally-symmetric tensors of even rank from the $Sp(2n)$ point of view. As we have mentioned, in the theory with fermions $f_\alpha$ one can find operators with more complicated types of symmetry. It should be possible to generalize the classification of the $Sp(2n)$-invariant correlators to the case of fields with mixed-symmetry as well as to consider the hyperfields which are $p$-forms in
hyperspace (see \cite{Alba:2015upa} for a discussion of $p$-forms
in $SO(2,D)$ CFTs).

Another interesting application of the generalized conformal $Sp(2n)$ symmetry is the study of conformal higher-spin fields on $Sp(n)$ group manifolds
 (see \cite{Didenko:2003aa,Plyushchay:2003gv, Plyushchay:2003tj,
Florakis:2014kfa, Florakis:2014aaa}
for details), 
which is a generalisation of conformal higher-spin theories on $AdS_D$
backgrounds (see for example \cite{Tseytlin:2013jya,Metsaev:2014iwa, Metsaev:2014sfa}).
For instance, the infinite sets of bosonic and fermionic symmetric higher-spin fields in $AdS_4$ are packed into a scalar and a spinor field which propagate on a 10-dimensional group manifold $Sp(4)$ 
and enjoy $Sp(8)$ invariance. As was shown in \cite{Florakis:2014kfa, Florakis:2014aaa} the correlation functions of the fields on $Sp(4)$ can be obtained from the flat hyperspace ones by performing a certain $GL(4)$ transformation and rescaling of the latter. So the results of this paper are directly generalized to $Sp(2n)$-invariant systems on the $Sp(n)$ group manifolds.

One more application of the free $Sp(2n)$-invariant systems is to compute partition functions of free higher-spin theories along the lines of \cite{Beccaria:2014jxa,Giombi:2014yra} (and references therein), an advantage being that $Sp(2n)$-fields encode the infinite multiplets of higher-spin fields and therefore should evaluate the sum over the spins automatically.

\subsection*{\bf Acknowledgments}
We are grateful to P. Dempster for his collaboration on a preliminary stage of this project and
to A. Barvinsky,  V. Didenko, M. Grigoriev and M. Vasiliev for fruitful discussions. The work of E.S. and D.S. was supported by the Russian Science Foundation grant 14-42-00047 in association with Lebedev Physical Institute. The work of E.S. was
supported by the DFG Transregional Collaborative Research Centre TRR 33. The work of D.S. and M.T. was supported by the Australian Research Council grant DP160103633. E.S. and D.S. also acknowledge a kind hospitality extended to them at the program ``Higher Spin Theory and Duality" MIAPP, Munich (May 2-27, 2016) organized by the Munich Institute for Astro- and Particle Physics (MIAPP) of the DFG cluster of excellence "Origin and Structure of the Universe".

\setcounter{equation}0
\appendix
\numberwithin{equation}{section}

\section{Properties of  $P$ and $Q$ tensor structures}\label{Appendix B}
The derivatives of the matrix valued coordinates
and corresponding determinants have the form
\begin{align} \label{xder}
\frac{\partial}{ \partial X^{\mu \nu}} X^{\ga\gb} &=
\frac12
\left(\delta^{\ga}_\mu \delta^{\gb}_\nu +
\delta^{\gb}_\mu \delta^{\ga}_\nu \right) \\
\frac{\partial}{ \partial X^{\mu \nu}} X^{-1}_{\ga\gb} &=-\frac12 \left(X_{\mu\alpha}^{-1} X_{\nu\beta}^{-1}+X_{\nu\alpha}^{-1} X_{\mu\beta}^{-1}\right)\\
    \frac{\partial}{ \partial X^{\mu \nu}} \det X &= X^{-1}_{\mu\nu} \det X
\end{align}
These relations can be used to derive 
useful properties of
the tensors $(P_{ab})_{\alpha \beta, \mu \nu}$ and $(Q^c_{ab})_{\alpha \beta}$, in particular
\begin{equation}
\frac{\partial}{ \partial X_3^{\mu \nu}} (Q^{3}_{12})_{\alpha \beta}=
\frac{1}{2} ( (P_{13})_{\alpha \beta, \mu \nu} - (P_{23})_{\alpha \beta, \mu \nu})
\end{equation}
\begin{equation}
\frac{\partial}{ \partial X_2^{\mu \nu}} (Q^{3}_{12})_{\alpha \beta}=
\frac{1}{2} (P_{23})_{\alpha \beta, \mu \nu}
\end{equation}
\begin{equation}
\frac{\partial}{ \partial X_1^{\mu \nu}} (Q^{3}_{12})_{\alpha \beta}=-
\frac{1}{2} (P_{13})_{\alpha \beta, \mu \nu}
\end{equation}
\begin{eqnarray}
\frac{\partial}{ \partial X_1^{\gamma \delta}} (P_{12})_{\alpha \beta, \mu \nu}&=& 
-\frac{1}{2} ( (P_{12})_{\mu \alpha, \gamma \delta} (X^{-1}_{12})_{\nu \beta} +
(P_{12})_{\nu \beta, \gamma \delta} (X^{-1}_{12})_{\mu \alpha}+ \\ \nonumber
&+&(P_{12})_{\nu \alpha, \gamma \delta} (X^{-1}_{12})_{\mu \beta}+
(P_{12})_{\mu \beta, \gamma \delta} (X^{-1}_{12})_{\nu \alpha}) \nonumber
\end{eqnarray}

The tensors $P$ and $Q$ have
the following properties under the differentiation which defines the conservation laws \eqref{conservation-j} and \eqref{conservation}
\begin{eqnarray} \label{consq1}
&&\frac{\partial}{ \partial X_3^{\mu \nu}} (Q^{3}_{12})_{\alpha \beta}-
\frac{\partial}{ \partial X_3^{\mu \alpha}} (Q^{3}_{12})_{\nu \beta}-
\frac{\partial}{ \partial X_3^{\nu \beta}} (Q^{3}_{12})_{\alpha \mu{}}+
\frac{\partial}{ \partial X_3^{\alpha \beta}} (Q^{3}_{12})_{\mu \nu}
= \\ \nonumber
&&=(P_{13})_{\alpha \beta, \mu \nu} - (P_{13})_{\alpha \mu, \beta \nu}
-(P_{23})_{\alpha \beta, \mu \nu} + (P_{23})_{\alpha \mu, \beta \nu}
\end{eqnarray}
\begin{eqnarray} \label{consq2}
&&\frac{\partial}{ \partial X_2^{\mu \nu}} (Q^{3}_{12})_{\alpha \beta}-
\frac{\partial}{ \partial X_2^{\mu \alpha}} (Q^{3}_{12})_{\nu \beta}-
\frac{\partial}{ \partial X_2^{\nu \beta}} (Q^{3}_{12})_{\alpha \mu{}}+
\frac{\partial}{ \partial X_2^{\alpha \beta}} (Q^{3}_{12})_{\mu \nu}
= \\ \nonumber
&&=(P_{23})_{\alpha \beta, \mu \nu} - (P_{23})_{\alpha \mu, \beta \nu}
\end{eqnarray}
\begin{eqnarray} \label{consq3} \nonumber
&&\frac{\partial}{ \partial X_1^{\mu \nu}} (Q^{3}_{12})_{\alpha \beta}-
\frac{\partial}{ \partial X_1^{\mu \alpha}} (Q^{3}_{12})_{\nu \beta}-
\frac{\partial}{ \partial X_1^{\nu \beta}} (Q^{3}_{12})_{\alpha \mu{}}+
\frac{\partial}{ \partial X_1^{\alpha \beta}} (Q^{3}_{12})_{\mu \nu}
= \\ 
&&=-(P_{13})_{\alpha \beta, \mu \nu} + (P_{13})_{\alpha \mu, \beta \nu}\,,
\end{eqnarray}
and finally
\begin{eqnarray} \label{consp} \nonumber
&&\frac{\partial}{ \partial X_1^{\gamma \delta}} (P_{12})_{\alpha \beta, \mu \nu}-
\frac{\partial}{ \partial X_1^{\gamma \alpha}} (P_{12})_{\delta \beta, \mu \nu}-
\frac{\partial}{ \partial X_1^{\beta \delta}} (P_{12})_{\alpha \gamma, \mu \nu}+
\frac{\partial}{ \partial X_1^{\alpha \beta}} (P_{12})_{\gamma \delta, \mu \nu}=
\\ 
&&=-(X^{-1}_{12})_{\alpha \gamma}(P_{12})_{\beta \delta, \mu \nu} -
(X^{-1}_{12})_{\beta \delta}(P_{12})_{\alpha \gamma , \mu \nu}+ \\ \nonumber
&&+(X^{-1}_{12})_{\alpha \beta}(P_{12})_{\gamma \delta, \mu \nu}+
(X^{-1}_{12})_{\gamma \delta}(P_{12})_{\alpha \beta, \mu \nu}\,.
\end{eqnarray}

\section{Conservation of $\langle JJ\mathcal O\rangle$ in detail}\label{Appendix C}

In this Appendix we present the calculations of the
current conservation in the three-point function
\p{jjo}. Below we omit $O({ N})$   indices, since they are not relevant for our goal.

First let us introduce the following notation
\begin{equation}\label{XYZ}
X_{\mu \nu} = (X_{13})^{-1}_{ \mu \nu}, \quad  Y_{\mu \nu} =
(X_{23})^{-1}_{ \mu \nu}, \quad Z_{\mu \nu} = (X_{12})^{-1}_{ \mu \nu}
\end{equation}
Therefore
\begin{equation}
({Q^3}_{12})_{\alpha \beta} = X_{\alpha \beta}-Y_{\alpha \beta},
\quad
(Q^1_{23})_{{\hat \mu} {\hat \nu}}= -Z_{{\hat \mu} {\hat \nu}}+
X_{{\hat \mu} {\hat \nu}}
\end{equation}
and
\begin{equation}
 (P_{13})_{\alpha \beta, \hat \mu \hat \nu}
 =
(X_{\alpha \hat \mu} X_{ \beta \hat \nu}
+   X_{\alpha \hat \nu} X_{ \beta \hat \mu})
\end{equation}
For the derivatives with respect to the coordinate
$X_3^{\mu \nu}$ one obtains
\begin{equation}
\partial_{\mu \nu}^{(3)}
({Q^3}_{12})_{\alpha \beta}
=\frac{1}{2}
(X_{\alpha \mu} X_{ \beta \nu}
+   X_{\alpha \nu} X_{ \beta \mu}
- Y_{\alpha \mu} Y_{ \beta \nu}
- Y_{\alpha \nu} Y_{ \beta \mu})
\end{equation}
\begin{equation}
\partial_{\mu \nu}^{(3)}
(Q^1_{23})_{{\hat \mu} {\hat \nu}}
 =\frac{1}{2}
(
X_{{\hat \mu} \mu} X_{ {\hat \nu} \nu}
+ X_{{\hat \mu} \nu} X_{ {\hat \nu} \mu})
\end{equation}
Now let us impose the conservation of the current
$J_{\alpha \beta}(X_3)$ on the three--point function
\begin{eqnarray}
&&G_{\alpha \beta {\hat \mu} {\hat \nu}} =
\langle   J_{{\hat \mu} {\hat \nu}}(X_1)
{\mathcal O}(X_2) J_{\alpha \beta } (X_3)
   \rangle
= \\ \nonumber
&&=(\det Z)^{-k_3} (\det X)^{-k_2} (\det Y)^{-k_1}
\left (  {\cal A} ({Q^3}_{12})_{\alpha \beta}
(Q^1_{23})_{{\hat \mu} {\hat \nu}}+
{\cal B} (P_{13})_{\alpha \beta, {\hat \mu} {\hat \nu} }   \right )
\end{eqnarray}
which explicitly reads
\begin{equation}\label{clo}
\partial_{\mu \nu}^{(3)} G_{\alpha \beta {\hat \mu} {\hat \nu}}-
\partial_{\mu \alpha}^{(3)} G_{\nu \beta {\hat \mu} {\hat \nu}}
-\partial_{\beta \nu}^{(3)} G_{\alpha \mu {\hat \mu} {\hat \nu}}
+ \partial_{\alpha \beta}^{(3)} G_{\mu \nu {\hat \mu} {\hat \nu}}=0,
\end{equation}
where
\begin{eqnarray}\label{dG}
&&\partial_{\mu \nu}^{(3)} G_{\alpha \beta {\hat \mu} {\hat \nu}}=
(\det Z)^{-k_3} (\det X)^{-k_2} (\det Y)^{-k_1} \times
\\ \nonumber
&&\times [{\cal A} ( \partial_{\mu \nu}^{(3)}
({Q^3}_{12})_{\alpha \beta}) (Q^1_{23})_{{\hat \mu} {\hat \nu}}+
{\cal A}  ({Q^3}_{12})_{\alpha \beta} (\partial_{\mu \nu}^{(3)}
(Q^1_{23})_{{\hat \mu} {\hat \nu}})
 +{\cal B}
(\partial_{\mu \nu}^{(3)} (P_{12})_{\alpha \beta, {\hat \mu} {\hat \nu}} ) + \\ \nonumber
&& +(k_2 X_{\mu \nu} + k_1 Y_{\mu \nu}) (
{\cal A} ({Q^3}_{12})_{\alpha \beta}
(Q^1_{23})_{{\hat \mu} {\hat \nu}}+
{\cal B} (P_{13})_{\alpha \beta, {\hat \mu} {\hat \nu} }
 )]
\end{eqnarray}
The other three terms in the conservation
law \p{clo} can be obtained from \p{dG} via interchange of the   indices.

\subsection{$\partial_{\mu \nu}^{(3)} G_{\alpha \beta {\hat \mu} {\hat \nu}}$}
The expressions  which are present in
\p{dG} have the following explicit form
\begin{eqnarray}\label{1}
&&  ( \partial_{\mu \nu}^{(3)}
({Q^3}_{12})_{\alpha \beta} ) (Q^1_{23})_{{\hat \mu} {\hat \nu}} = \\ \nonumber
&&=+\frac{1}{2 } [ X_{\alpha \mu} X_{\beta \nu} X_{{\hat \mu} {\hat \nu}}
+ X_{\alpha \nu} X_{\beta \mu} X_{{\hat \mu} {\hat \nu}}
- Y_{\alpha \mu} Y_{\beta \nu} X_{{\hat \mu} {\hat \nu}}
- Y_{\alpha \nu} Y_{\beta \mu} X_{{\hat \mu} {\hat \nu}} ] -\\ \nonumber
&& -\frac{1}{2 } [ X_{\alpha \mu} X_{\beta \nu} Z_{{\hat \mu} {\hat \nu}}
+ X_{\alpha \nu} X_{\beta \mu} Z_{{\hat \mu} {\hat \nu}}
- Y_{\alpha \mu} Y_{\beta \nu} Z_{{\hat \mu} {\hat \nu}}
- Y_{\alpha \nu} Y_{\beta \mu} Z_{{\hat \mu} {\hat \nu}} ]
\end{eqnarray}

\begin{eqnarray}
&&   ({Q^3}_{12})_{\alpha \beta} (\partial_{\mu \nu}^{(3)}
(Q^1_{23})_{{\hat \mu} {\hat \nu}}) = \\ \nonumber
&&=\frac{1}{2 } [ X_{{\hat \mu} \mu} X_{{\hat \nu} \nu} X_{\alpha \beta}
+X_{{\hat \mu} \nu} X_{{\hat \nu} \mu} X_{\alpha \beta}
-X_{{\hat \mu} \mu} X_{{\hat \nu} \nu} Y_{\alpha \beta}
-X_{{\hat \mu} \nu} X_{{\hat \nu} \mu} Y_{\alpha \beta}
]
\end{eqnarray}

\begin{eqnarray}
&& \partial_{\mu \nu}^{(3)}
(P_{13})_{\alpha \beta, {\hat \mu} {\hat \nu}}  = \\ \nonumber
&&=\frac{1}{2 } [ X_{\alpha \mu} X_{{\hat \mu} \nu} X_{\beta {\hat \nu}}
+X_{\alpha \nu} X_{{\hat \mu} \mu} X_{\beta {\hat \nu}}
+X_{\alpha {\hat \mu}} X_{\beta \mu} X_{ {\hat \nu} \nu}
+X_{\alpha {\hat \mu}} X_{\beta \nu} X_{ {\hat \nu} \mu}
 ] +\\ \nonumber
&& + \frac{1}{2 } [ X_{\alpha \mu} X_{{\hat \nu} \nu} X_{\beta {\hat \mu}}
+ X_{\alpha \nu} X_{{\hat \nu} \mu} X_{\beta {\hat \mu}}
+X_{\alpha {\hat \nu}} X_{\beta \mu} X_{ {\hat \mu} \nu}
+X_{\alpha {\hat \nu}} X_{\beta \nu} X_{ {\hat \mu} \mu}]
\end{eqnarray}

\begin{eqnarray}
&&(k_2 X_{\mu \nu} + k_1 Y_{\mu \nu})
({Q^3}_{12})_{\alpha \beta}  (Q^1_{23})_{{\hat \mu} {\hat \nu}}
= \\ \nonumber
&&=k_2[X_{\mu \nu} X_{\alpha \beta} X_{{\hat \mu} {\hat \nu}}
- X_{\mu \nu} X_{\alpha \beta} Z_{{\hat \mu} {\hat \nu}}
- X_{\mu \nu}Y_{\alpha \beta} X_{{\hat \mu} {\hat \nu}}
+ X_{\mu \nu}Y_{\alpha \beta}Z_{{\hat \mu} {\hat \nu}}]+ \\ \nonumber
&&+k_1[Y_{\mu \nu} X_{\alpha \beta} X_{{\hat \mu} {\hat \nu}}
- Y_{\mu \nu} X_{\alpha \beta} Z_{{\hat \mu} {\hat \nu}}
- Y_{\mu \nu}Y_{\alpha \beta} X_{{\hat \mu} {\hat \nu}}
+ Y_{\mu \nu}Y_{\alpha \beta}Z_{{\hat \mu} {\hat \nu}}]
\end{eqnarray}

\begin{eqnarray}\label{5}
&&(k_2 X_{\mu \nu} + k_1 Y_{\mu \nu})
(P_{13})_{\alpha \beta, {\hat \mu} {\hat \nu}}= \\ \nonumber
&&=k_2[X_{\mu \nu} X_{\alpha {\hat \mu}} X_{\beta {\hat \nu}}
+ X_{\mu \nu} X_{\alpha {\hat \nu}} X_{\beta {\hat \mu}}
] +\\ \nonumber
&&+k_1[Y_{\mu \nu} X_{\alpha {\hat \mu}} X_{\beta {\hat \nu}}
+ Y_{\mu \nu} X_{\alpha {\hat \nu}} X_{\beta {\hat \mu}}]
\end{eqnarray}
Again the terms in
$\partial_{\mu \alpha}^{(3)} G_{\nu \beta {\hat \mu} {\hat \nu}}$,
$\partial_{\beta \nu}^{(3)}
G_{\alpha \mu {\hat \mu} {\hat \nu}}$ and
$ \partial_{\alpha \beta}^{(3)} G_{\mu \nu {\hat \mu} {\hat \nu}}$
can be obtained from \p{1}--\p{5} via appropriate
interchanging of the   indices. After doing so and collecting similar terms one
obtains the equations \p{usl1}. Obviously the same condition \p{usl1} can be obtained if one considers
the conservation of the current $J_{\hat \mu \hat \nu}(X_1)$ instead of $J_{\alpha \beta}(X_3)$.

\subsection{Three Dimensions}
In this subsection we shall explicitly show that in the case $n=2$, $D=3$ win which the generalised conformal group $Sp(4)$ coincides with the three--dimensional conformal group,
the current conservation condition in the three point function with two currents and one scalar does not impose
any restriction on the scaling dimension of the later, thus reproducing a known result from $D=3$ CFT (see for example
\cite{Giombi:2011rz}).

In $D=3$ the coordinates $X^{\alpha \beta}=x^m\gamma_m^{\alpha\beta}$ $(m=0,1,2)$
are $2\times 2$ symmetric matrices parametrising the $D=3$ space-time. 
Taking the generalized conformal dimensions of the current to be equal to $1$ we obtain from \p{3k-1} that
  $k_2=1-k_1$.
The conventional conservation law is obtained by contracting expressions \eqref{1}--\eqref{5} with $\epsilon^{\alpha\nu}\epsilon^{\beta\mu}$, 
where the convention for the spinorial metric is
$
\epsilon^{\hat\mu \hat\nu}\epsilon_{\hat\nu \hat\rho}=-\delta^{\hat\mu}_{\hat\rho}.
$
 Doing so one gets
\begin{eqnarray}\label{sum}
&&-\frac {\cal A}2(X^2-Y^2)(X_{\hat \mu \hat \nu}-Z_{\hat \mu \hat \nu})+
{\cal A}[X(X-Y)X]_{\hat\mu \hat\nu} -  \\ \nonumber
&&-
2{\cal B}[XXX]_{\hat\mu \hat \nu}+A(X^2-XY)(X_{\hat \mu \hat \nu}-Z_{\hat\mu \hat\nu})-\\ \nonumber
&&-{\cal A}k_1 (X^2-2XY+Y^2)(X_{\hat\mu \hat\nu}-Z_{\hat\mu \hat\nu}) +2
{\cal B}[XXX]_{\hat\mu \hat\nu}-2Bk_1[X(X-Y)X]_{\hat\mu \hat\nu}=\\ \nonumber
&&={\cal A}(\frac 12-k_1)(X-Y)^2(X_{\hat\mu \hat\nu}-Z_{\hat\mu \hat\nu})+
({\cal A}-2{\cal B}k_1)[X(X-Y)X]_{\hat\mu \hat\nu}=0\,,\\ \nonumber
\end{eqnarray}
where
$$
XY=X_{\alpha\beta}Y^{\alpha\beta}, \qquad [X(X-Y)X]_{\hat\nu \hat\mu}=[X(X-Y)X]_{\hat\mu \hat\nu}=X_{\hat\mu}{}^\alpha(X-Y)_{\alpha\beta}
X^{\beta}{}_{\hat \nu}.
$$
For the expression \eqref{sum} to be zero for an arbitrary $k_1$ one should prove that
\begin{equation}\label{12}
(X-Y)^2(X_{\hat\mu \hat\nu}-Z_{\hat\mu \hat\nu})=a [X(X-Y)X]_{\hat\mu \hat\nu}\,.
\end{equation}
To this end note that in vew of the definitions \eqref{XYZ} the following relation holds
$$
Z^{-1}=X^{-1}-Y^{-1}.
$$
Note also that for the $2\times 2$ symmetric matrices $X_{\mu\nu}$
$$
X^2=2 \det X=\frac {2} {\det X^{-1}}\,.
$$
Now using that
$$
X_{\hat\mu}{}^{ \beta} X_{ \beta  \hat\nu}=-\frac 12 \epsilon_{\hat\mu \hat\nu} X^2,\qquad   \rightarrow \qquad X_{\hat\mu \hat\nu}=\frac 12 X^2 X^{-1}_{\hat\mu \hat\nu} = 
\frac 1{\det X^{-1}} X^{-1}_{\hat\mu \hat\nu}\,,
$$
we can rewrite $[X(X-Y)X]_{nm}$ as follows
\bea\label{XXX}
X_{\hat\mu}{}^\alpha(X-Y)_{\alpha\beta}X^{\beta}{}_{\hat \nu}&=&\frac 12 X^2X_{\hat\mu \hat\nu}+\frac 12 X^2Y_{\hat\mu \hat\nu}-(XY)
X_{\hat\mu \hat\nu}\\
&=&\frac 12 (X-Y)^2X_{\hat\mu \hat\nu}-\frac 12 Y^2X_{\hat\mu \hat\nu}+\frac 12 X^2Y_{\hat\mu \hat\nu}\nonumber\\
&=&\frac 12 (X-Y)^2X_{\hat\mu \hat\nu}-\frac 14 X^2Y^2(X^{-1}-Y^{-1})_{\hat\mu \hat\nu} \nonumber\\
&=&\frac 12 (X-Y)^2X_{\hat\mu \hat\nu}-\frac 14 X^2Y^2\det(X^{-1}-Y^{-1})\frac{(X^{-1}-Y^{-1})_{\hat\mu \hat\nu}}{\det(X^{-1}-Y^{-1})}\nonumber\\
&=&\frac 12 (X-Y)^2X_{\hat\mu \hat\nu}- \det X\det Y\det(X^{-1}-Y^{-1}) Z_{\hat\mu \hat\nu}\nonumber\\
&=&\frac 12 (X-Y)^2X_{\hat\mu \hat\nu}-\det(X-Y) Z_{\hat\mu \hat\nu} \nonumber \\
&=&\frac 12 (X-Y)^2(X_{\hat\mu \hat\nu}-Z_{\hat\mu \hat\nu})\,.\nonumber
\eea
We thus find that in \eqref{12}  $a=2$. Hence, from \eqref{sum} it follows that
$$
{\cal A}(1-k_1)-{\cal B}k_1=0,
$$
which means that in this case there is no restriction on the parameter $k_1$ and hence on the conformal dimension of the scalar field.

\if{}
 \bibliographystyle{utphys}
 \bibliography{references}

\end{document}
\fi

\providecommand{\href}[2]{#2}\begingroup\raggedright\endgroup

\end{document}